\documentclass[12pt]{amsart}

\usepackage[T1]{fontenc}
\usepackage[margin=1.2in]{geometry}
\usepackage{amsaddr,amsmath,amssymb,amsthm,mathtools}
\usepackage{microtype}
\usepackage[normalem]{ulem}
\usepackage[authoryear,square]{natbib}
\usepackage{verbatim}

\usepackage[colorlinks=true,linkcolor=blue,citecolor=blue,urlcolor=blue]{hyperref}
\usepackage{enumitem}
\setlist[enumerate]{itemsep=1pt,topsep=3pt}

\usepackage{setspace}
\makeatletter
\renewcommand{\section}{\@startsection{section}{1}%
  \z@%
  {1.5\linespacing\@plus.5\linespacing}%
  {.5\linespacing}%
  {\normalfont\scshape\centering}}
\makeatother

\newtheorem{lemma}{Lemma}
\newtheorem{theorem}[lemma]{Theorem}
\newtheorem{corollary}[lemma]{Corollary}
\newtheorem{proposition}[lemma]{Proposition}
\newtheorem{definition}[lemma]{Definition}

\theoremstyle{remark}

\def\Re{\mathbf{R}} \def\R{\mathbf{R}}
 
\def\Z{\mathbf{Z}}

\def\R{\mathbb R}
\def\Re{\mathbb R}
\def\Z{\mathbb Z}

\def\phi{\varphi}

\def\one{\mathbf{1}}

\def\argmax{\mbox{argmax}} \def\argmin{\mbox{argmin}}

\DeclareMathOperator{\dist}{dist}
\renewcommand{\argmax}{\operatorname*{arg\,max}}
\renewcommand{\argmin}{\operatorname*{arg\,min}}

\DeclareMathOperator{\conv}{conv}
\DeclareMathOperator{\cone}{cone}

\makeatletter
\newcommand{\defterm}[1]{\@defterm#1 \@empty\@nil}
\def\@defterm#1 #2\@nil{%
  \uline{#1}%
  \ifx\@empty#2\@empty
  \else
    \space\@defterm#2\@nil
  \fi
}
\makeatother

\newcommand{\df}[1]{\textit{\defterm{#1}}}

\newcommand{\ip}[2]{\langle #1,#2\rangle}
\newcommand{\norm}[1]{\lVert #1\rVert}
\newcommand{\pos}[1]{[#1]_+}

\def\bun{\mathcal{S}}

\title[Combinatorial assignment]{Fairness and incentives in combinatorial assignment}

\author{Federico Echenique}
\address{Department of Economics, UC Berkeley}
\author{Albin Erlanson}
\address{Department of Economics, University of Essex}
\author{Anna Vakarova}
\address{Department of Economics, UC Berkeley}

\begin{document}

\begin{abstract}
We consider the fair and efficient random assignment of indivisible goods subject to resource constraints. Agents receive bundles randomly, but individual lotteries over bundles that are feasible in expectation need not be jointly feasible. Consequently, we treat a lottery over feasible allocations as a public good. We show that every feasible finite economy admits an envy-free and Pareto efficient random assignment that may be supported as a Lindahl equilibrium. For a finite type space with specified expected utilities, we construct a mechanism  whose outcomes are exactly feasible and Pareto efficient with
respect to reported preferences. As the population grows and empirical type proportions and per-capita resources approach a limiting large economy, envy and gains from manipulation vanish. The approximate incentive compatibility of the mechanism holds for all type profiles that approximate the limiting economy. 
The argument combines sensitivity bounds for linear programs with proximity bounds between fractional and integer optima.
\end{abstract}

\maketitle

\clearpage
\thispagestyle{empty}

\tableofcontents
\clearpage

\setcounter{page}{1}
\section{Introduction}

We consider the fair and efficient assignment of indivisible goods to a collection of agents and the agents' incentives to report their preferences. Agents receive bundles of goods subject to resource constraints, and indivisibilities often preclude the existence of a fair deterministic assignment. Thus, we have to allow random assignments.  Our fairness requirement is that no agent would prefer another agent's lottery over bundles to her own, so called envy-freeness. Efficiency is also evaluated before the randomness is resolved, using agents' expected utilities. We prove the existence of fair and efficient random assignments; we show that these can be decentralized as Lindahl equilibria, and study the problem of incentive compatible preference reporting needed to calculate the assignment.

In the approach of \citet{HyllandZeckhauser1979}, agents buy lotteries over objects in a pseudomarket. When their incomes are equal, all agents face the same budget set and, consequently, cannot envy another agent's lottery. A version of the first welfare theorem ensures that the assignment is efficient. In their setting, the Birkhoff--von Neumann theorem provides the connection between individual lotteries and feasible assignments: every collection of marginal lotteries satisfying the capacity constraints in expectation can be implemented through a lottery over feasible deterministic assignments. Feasibility in expectation is therefore enough to ensure that every realization of a random assignment is feasible. 

The Birkhoff--von Neumann property can fail when assignments involve several goods or dates. Consider, for example, the assignment of dorm rooms to college students over four years. A student's bundle is a four-year sequence of rooms, and her preferences may depend on the complete sequence. Even if individual lotteries respect every room's capacity in every year in expectation, they need not be jointly implementable. To see this, consider two students and two rooms, $A$ and $B$, available in each of the four years. The first student is offered an equal chance of spending all four years in $A$ or all four years in $B$. The second is offered an equal chance of the sequences $(A,B,A,B)$ and $(B,A,B,A)$. Each room is occupied by one student in expectation in every year. But these lotteries cannot be implemented together: whichever room the first student receives for all four years, the second student must receive the other room for all four years. Other examples where the Birkhoff--von Neumann property fails include general combinatorial assignments, such as the problem of course bidding (\cite{BudishEtAl2013} provide general conditions that ensure the Birkhoff--von Neumann property). 

To deal with these restrictions, we treat a random assignment as a public good. The collective choice is a single lottery over feasible economy-wide allocations. All agents evaluate this common choice of a random assignment, even if each individual only cares about their private consumption. We prove that every feasible finite economy admits an envy-free and Pareto efficient random assignment. We also show that every Pareto efficient random assignment can be decentralized as a Lindahl equilibrium, a version of the second welfare theorem for our model. Personalized Lindahl prices support the choice of a common allocation while allowing each individual agent to value it differently.

Last but not least, we are concerned with agents' incentives to report their preferences, which determine how they evaluate alternative assignments. We assume a finite type space and a given utility function over bundles for every type. An agent's type is private information, and lotteries are evaluated according to expected utility. We fix a reference large economy, specifying a positive mass of every type and a vector of per-capita resources. Think of a measure space of agents with a non-atomic continuum of agents of each type. From this economy, we select an envy-free and efficient target assignment, together with strictly positive welfare weights supporting it. Our mechanism uses these fixed weights to maximize weighted utilitarian welfare in a finite economy and chooses the implementable mass vector closest to the target among the welfare maximizers.

For every finite economy, the mechanism assigns a lottery over feasible deterministic allocations that is Pareto efficient with respect to the reported types. When the population is large and the empirical type distribution and per-capita resources are close to those of the reference economy, the mechanism is also approximately envy-free and approximately incentive compatible. The limiting large-market economy is an input into the definition of the mechanism, an assumption that is analogous to a known prior in Bayesian mechanism design; however, our incentive guarantee does not require a common prior or equilibrium behavior by the other agents. It is stronger than that. Fix an agent and arbitrary reports by her opponents. Truthful reporting is an approximately best response whenever those reports, together with her truthful report, describe an economy sufficiently close to the limiting economy.

The intuition for approximate incentive compatibility comes from the nonatomic limiting economy. In the limiting economy, individual agents are negligible and  cannot change the menu of lotteries assigned to the different types. Envy-freeness then ensures that she would not prefer the lottery intended for another type. In a finite economy, however, a false report changes both the agent's reported type and the allocation problem that the mechanism solves. The argument for approximate incentive compatibility therefore requires controlling the consequences of this change.

While intuitive, the argument for approximate incentive compatibility is subtle. The target assignment and welfare weights remain fixed while the finite economy grows. As a finite economy grows and approaches the limiting large economy, the set of feasible welfare-maximizing allocations changes. We deal with this approximation using new techniques borrowed from sensitivity analysis in the optimization literature. These techniques control how the fractional optimal sets vary with the parameters of the finite economy as these approximate the limiting economy. Integer proximity bounds then relate fractional optima to allocations that are  implementable in the finite economy. Together, these estimates show that the lotteries offered by the mechanism remain close to the reference lotteries, both before and after a unilateral false report. Envy-freeness of the target assignment in the limiting economy then bounds the gain from manipulation.

\section{Related Literature}

Our existence result is related to the literature on fair and efficient allocations. The early work of \citet{Foley1967} and \citet{Varian1974} develops envy-freeness as a criterion for evaluating allocations and studies its compatibility with Pareto efficiency. Their models assume infinitely divisible goods and monotonic preferences: a competitive equilibrium with equal incomes is then fair and efficient. \citet{HyllandZeckhauser1979} bring this approach to the allocation of indivisible objects under a unit demand constraint (each agent is allocated at most one good). Their pseudomarket allocates probability shares in objects, which are subsequently implemented through a lottery over assignments that is made possible by the Birkhoff von-Neumann theorem. Overall, in these papers, equal budgets give agents the same opportunities at equilibrium prices: each agent can afford every other agent's consumption and nonetheless chooses her own. A version of the first-welfare theorem implies that equilibrium allocations are also Pareto optimal. In the same discrete-good, unit-demand model as \citeauthor{HyllandZeckhauser1979}, \citet{BogomolnaiaMoulin2001} study fair and efficient random assignment using only ordinal preference information. Their probabilistic serial mechanism is envy-free in the sense of stochastic dominance and is ordinally efficient. Our result generalizes beyond the unit demand setting (indeed, to models where no analogous result to Birkhoff-von Neumann is available); it uses fixed von Neumann–Morgenstern utilities and establishes ex-ante Pareto efficiency with respect to the associated expected utilities. 

A general existence result is due to \cite{vakarova26}, who shows that a weak Pareto optimal and envy-free allocation exists in very general economic environments, including those under the assumptions in our model. Like us, her existence argument relies on finding welfare-weights that, in a weighted utilitarian optimization, result in an allocation with the desirable properties.

The closest paper to ours is \cite{citanna2026envy}, which also draws a connection between fair and efficient random allocations, Lindahl equilibria, and approximate incentive compatibility. Their work is simultaneous and independent of ours. Citanna and Tvede consider a model with endowments and individual admissibility restrictions and focus on a notion of fairness based on net trades. They study mechanisms that achieve fairness, constrained Pareto optimality (constrained to fair outcomes), and approximate incentive compatibility. Separately, they present an existence result for (full) efficiency and fairness. Their incentive compatibility argument relies on the generic uniqueness of optimal marginal allocations (as in \cite{Jackson1992}), and diffuse individual welfare weights. Our argument relies on different techniques. We allow for multiple optima and use the stability of optimal sets and integral approximation to obtain explicit bounds on how far the agents' marginal lotteries are from those in the reference economy and on the effect of an individual report on the aggregate outcome. The former allows us to derive explicit bounds on envy and the gain from manipulation.

The work of \cite{echenique21} also deals with general allocation problems in which Birkhoff--von-Neumann may fail, and may include individualized feasibility constraints. A solution involves pseudo market for individual lotteries and potentially personalized prices (but only personalized to the extent that feasibility constraints are personalized). \cite{ECHENIQUE2021105274} focuses on the unit demand model of \cite{HyllandZeckhauser1979}, but allows for endowments and property rights. 

A related approach for collective decision problems is developed by \citet{GulPesendorfer2026}, who study Lindahl equilibria in pseudo-markets with personalized prices for lotteries over social alternatives and characterize equilibrium payoffs through weighted Nash bargaining. Their general framework also encompasses combinatorial assignment problems.

Our results on approximate incentive compatibility mean that the incentives to manipulate a mechanism vanish in large economies. This is a long-standing idea in economic theory that originated with the problem of manipulability of Walrasian equilibria due to agents having market power, and the notion that this market power would vanish as the economy grows large \citep{Hurwicz1972}. We shall discuss a few of the most relevant precedents. Our incentive argument follows the approach of \citet{RobertsPostlewaite1976} and \citet{Jackson1992}: an individual's small effect on aggregate data leads to a small gain from manipulation when the limiting economy is sufficiently \textit{regular}. See also \citet{JacksonManelli1997}. The relevant property is the stability of market outcomes, which we achieve in our sensitivity analysis. \cite{Watt2026} is a recent example of this kind of result for Walrasian equilibrium in a model with quasi-linear utility. Our construction follows the reasoning in these foundational papers but obtains the analogous property to regularity from a fixed polyhedral allocation problem. Hoffman's bound controls changes in the fractional optimal set, while integer proximity transfers the comparison to exactly implementable finite allocations. 

\citet{Hammond1979} characterizes incentive-compatible mechanisms in a nonatomic economy through choice from menus that do not depend on the individual's own characteristics. This is analogous to envy-freeness in our work. In a finite economy, however, a manipulation changes the menus in question. We control this effect while requiring the lotteries to be jointly implementable as feasible deterministic allocations. \citet{KojimaManea2010} is a precedent in a model with random assignments. For a given expected utility function of an agent and a fixed set of object types, sufficiently many copies of every object type make truthful reporting under probabilistic serial a weakly dominant strategy for that agent, regardless of the other agents' preferences.
Their incentive conclusion is stronger than ours: it is eventually exact. Our model, instead, allows for complete bundles that use several resources, so elementary fractional feasibility need not imply joint implementability. We also require ex-ante Pareto optimality instead of ordinal efficiency.

\cite*{HeMirallesPyciaYan2018} study cardinal assignment problems without monetary transfers, including the housing model of discrete object allocations. They obtain incentive compatibility in large economies through an argument related to \citeauthor{Jackson1992}'~\citeyear{Jackson1992} regularity condition, which requires an analysis of  the behavior of the limiting equilibrium-price correspondence that is analogous to our sensitivity analysis of polyhedral optimal sets. While the standard unit-demand discrete allocation model is a special case, the focus in their paper is on the role of priorities, which is distinct from our model of assignment problems where the standard implementation (as in the Birkhoff von Neumann theorem) fails.

\citet{JacksonKremer2007} and \citet{AzevedoBudish2019} leverage envy-freeness for incentives in large economies. The basic argument is that when an individual has little influence on aggregate outcomes, reporting another type gives access to that type's allocation. Envy-freeness then rules out a gain from such a report. Our arguments follow the same logic. Jackson and Kremer identify limiting envy-freeness as a necessary condition for robust approximate implementation and establish sufficiency under additional assumptions. Azevedo and Budish show that an anonymous envy-free mechanism is strategy-proof in the large, evaluating incentives under every fixed full-support i.i.d.\ distribution of opponents' reports. In that sense, they obtain incentive compatibility in expectation rather than ex-post. In our case, knowledge of the target allocation in the limiting economy is key. Our result strengthens the incentive guarantee from interim expectations to realized profiles near the limiting economy, but narrows its scope by requiring advance specification and knowledge of the limiting economy.

We use random assignments to achieve fairness. The ex-ante notion of envy-freeness is in contrast to ex-post criteria such as envy-freeness up to one good, in which an agent's envy must disappear after removing a single good from the envied bundle \citep{Budish2011}. These criteria address different concerns. Random assignment allows for envy-freeness over prospects even when exact envy-freeness of realized allocations is impossible.

\section{The Model}\label{sec:model}

\subsection{Notation and Primitives}

The model primitives are a tuple $(\bun,\Theta,R,u)$, where
\begin{itemize}
    \item $\bun$ is a finite set of $m=|\bun|$ \df{bundles}.  
    \item $\Theta$ is a finite set of $q=|\Theta|\geq 2$ \df{types}.
    \item $R\in\Z_+^{k\times m}$ is a matrix of \df{resource uses}: column
    $R_s$ describes the resources used by bundle $s\in\bun$.
    \item $u=(u_\theta)_{\theta\in\Theta}$ is a collection of utility
    functions, with $u_\theta:\bun\to[0,M]$ and $M>0$.
\end{itemize}

A \df{finite economy} is a tuple $((\theta_i)_{i=1}^N,B)$, where $\{1,\dots,N\}$ is the set of agents, each of whom has a type $\theta_i\in \Theta$, and $B\in \Z_+^k$ is an integer \df{supply vector} representing the number of resources. Agents consume bundles in $\bun$. Their preferences over bundles depend on their type. In fact, we fix a von-Neumann-Morgenstern utility for each type that also captures their (fixed) preferences over random bundles. If an agent receives bundle $s\in\bun$, then they consume $R_{hs}$ units of resource $h$. The set of bundles and their resource uses are the same for all agents.

A \df{deterministic allocation} is a tuple $(s_1,\ldots,s_N)\in\bun^N$. A deterministic allocation is \df{feasible} if $\sum_{i=1}^N R_{s_i}=B$. We say that a pair $(N,B)$ is \df{feasible} if there exists a feasible deterministic allocation. Throughout our discussion of a finite economy, we assume that $(N,B)$ is feasible. For a fixed pair $(N,B)$, we may consider all type profiles $(\theta_i)_{i=1}^N$, and each one will, together with $B$, define a finite economy. The set of feasible allocations only depends on $(N,B)$, not on agents' types.

Our notion of feasibility differs from the modeling in most recent work on discrete allocation, but it is general enough to encompass a number of interesting applications (see Section~\ref{sec:applications}). In Section~\ref{sec:feasibility}, we present a partial characterization of the scope of our representation of feasible allocations.

A \df{random feasible allocation} is a probability measure over the set of feasible deterministic allocations. The set of all random feasible allocations is
\[
\Pi_{N,B}=\Delta\left(\left\{(s_1,\ldots,s_N)\in\bun^N \colon \sum_{i=1}^N R_{s_i}=B\right\}\right).
\]
Given $\pi\in\Pi_{N,B}$, let $(\tilde s_1,\ldots,\tilde s_N)$ denote the resulting random allocation. The \df{marginal random assignment} of agent $i$ is $x_i\in\Delta(\bun)$, where $x_{is}=\Pr_\pi(\tilde s_i=s)$. The expected utility of agent $i$ is
\[
\mathbb E_\pi[u_{\theta_i}(\tilde s_i)] =u_{\theta_i}\cdot x_i =\sum_{s\in\bun}u_{\theta_i}(s)x_{is}.
\]
If agents of the same type receive the same marginal assignment, we denote their common assignment by $x_\theta$ and say that the allocation
\df{treats identical types equally}.

\subsection{A large economy}

To describe a large economy, we specify a distribution $d\in\Delta^{q-1}$ of types and a vector $\sigma\in\Re_{++}^k$ of per-agent resources. Interpret $d_\theta$ as the mass of type-$\theta$ agents in an atomless population of total mass one.
Given the fixed primitives $(\bun,\Theta,R,u)$, a \df{large economy} is a tuple $(d,\sigma)$, with $d \gg 0$.

\subsection{Definitions}

\begin{definition}
  Let $\epsilon \geq 0$. We say that agent $i$ \df{$\epsilon$-envies} agent $j$ at $\pi$ if $u_{\theta_i}\cdot x_j > u_{\theta_i}\cdot x_i+\epsilon$.
Let $\pi\in\Pi_{N,B}$ be a random feasible allocation.
\begin{itemize}
    \item $\pi$ is \df{Pareto optimal} if there is no $\pi'\in\Pi_{N,B}$ such that
    \[
    \mathbb E_{\pi'}[u_{\theta_i}(\tilde s_i)]
    \geq\mathbb E_\pi[u_{\theta_i}(\tilde s_i)]
    \qquad(i=1,\ldots,N),
    \]
    with strict inequality for at least one agent.
    \item $\pi$ is \df{$\epsilon$-envy-free} if no agent $\epsilon$-envies another agent at $\pi$. It is \df{envy-free} if it is $0$-envy-free.
\end{itemize}
\end{definition}

Following \cite{HyllandZeckhauser1979}, the standard approach in the random allocation of discrete goods is to work with marginal assignments of goods to agents that satisfy the feasibility constraints in expectation. By the Birkhoff von-Neumann Theorem, these are always implementable by means of a randomization over feasible deterministic assignments.  In our setting, however, this is not true. Individual marginal assignments satisfying the resource constraints in expectation need not be implementable through a randomization over deterministic allocations. Our solution is two-fold. First, we work with an economy-wide allocation as a public good. We establish the existence of a fair and efficient allocation and show that it can be decentralized by means of a market solution. In line with the public good nature of our model, the decentralization involves a Lindahl equilibrium. Second, we keep careful track of the distinction between what can and cannot be implemented: see Lemma~\ref{lem:implementation} below. This becomes crucial as we present our large economy results.

Let $K$ be the number of feasible deterministic allocations. A personalized price for agent $i$ assigns a real number $p_i(s_1,\ldots,s_N)$ to each feasible deterministic allocation $(s_1,\ldots,s_N)$. Identifying these personalized prices with vectors in $\Re^K$, with one coordinate for each feasible deterministic allocation, we write $p_i\cdot\pi=\mathbb E_\pi[p_i(\tilde s_1,\ldots,\tilde s_N)]$.

\begin{samepage}
\begin{definition}\label{def:lindahl-finite}
A \df{Lindahl equilibrium} is a tuple $(\pi,p,w)$ in which
$\pi\in\Pi_{N,B}$ is a random feasible allocation,
$p=(p_1,\ldots,p_N)\in\Re^{KN}$ is a vector of
\df{personalized prices}, and $w\in\Re_+^N$ represents agents'
incomes, such that:
\begin{enumerate}
    \item For every agent $i$,
    \[
    \pi\in\argmax\left\{ \mathbb E_{\pi'}[u_{\theta_i}(\tilde s_i)]\colon p_i\cdot\pi'\leq w_i,\ \pi'\in\Pi_{N,B}\right\},
    \]
    and $\pi$ minimizes $p_i\cdot\pi'$ among all solutions to this maximization problem.
  \item
    \[
    \pi\in\argmax_{\pi'\in\Pi_{N,B}} \left(\sum_{i=1}^N p_i\right)\cdot\pi'.
    \]
\end{enumerate}
\end{definition}
\end{samepage}

\subsection{Integer count tables and mass vectors}

We focus on discrete assignment problems where an equivalence between marginal random assignments and random feasible allocations can fail. Here we build up the basic notions that allow us to keep track of these two kinds of objects. The result we need in our analysis is simple and is presented at the end of this section as Lemma~\ref{lem:implementation}. First, however, we need to build up the necessary definitions.

For any finite economy, let $n_\theta=|\{i:\theta_i=\theta\}|$ be the number of agents of type $\theta$. A deterministic allocation defines an integer \df{count table} $z=(z_{\theta s})$, where $z_{\theta s}$ is the number of type-$\theta$ agents receiving bundle $s$. A feasible count table satisfies
\[
\sum_{s\in\bun}z_{\theta s}=n_\theta \quad(\theta\in\Theta),\qquad \sum_{\theta\in\Theta}\sum_{s\in\bun}R_s z_{\theta s}=B.
\] Note that any feasible count table can be translated into a feasible deterministic allocation by assigning the agents bundles according to what their type is assigned in the count table. 

The feasibility conditions can be represented by an integer matrix $A$, with $q+k$ rows and $qm$ columns, whose $(\theta,s)$ column is
\[
A_{\theta s}=\begin{pmatrix}\one_\theta\\R_s\end{pmatrix}.
\]
Here $\one_\theta$ is the coordinate vector in $\Re^q$ whose $\kappa$ coordinate is $\one\{\kappa=\theta\}$. Normalizing the feasibility equations by $N$ preserves the linear feasibility conditions, allowing us to express the following variables in per-agent terms: 
\[
d_\theta=\frac{n_\theta}{N}, \qquad
d=(d_\theta)_{\theta\in\Theta}, \qquad \sigma=\frac BN,\quad \text{and} \quad b=(d,\sigma), 
\] and let 
\[
\mathcal Z_N(b) = \{z\in\Z_+^{qm}:Az=Nb\} 
\]
denote the set of feasible count tables. 

When $z\in \mathcal Z_N(b)$, the vector $z/N$ expresses a count table in per-agent terms and satisfies the linear system $A (z/N)=b$.  If we drop the restriction that the coordinates arise from an integer count table and allow any nonnegative real vector satisfying this linear system, we obtain a relaxation of $\frac{1}{N}\mathcal{Z}_N(b)$. These are called \df{mass vectors} and are collected in the following polytope, 
\begin{equation}\label{eqP} P(b)=\{y\in\Re_+^{qm}:Ay=b\}.
\end{equation}
Equivalently, mass vectors satisfy
\[
\sum_{s\in\bun}y_{\theta s}=d_\theta \quad(\theta\in\Theta),\qquad \sum_{\theta\in\Theta}\sum_{s\in\bun}R_s y_{\theta s}=\sigma.
\]
The normalized count tables are the vectors $z/N$ with $z\in\mathcal Z_N(b)$. Their convex hull is
\[
P_N(b)=\conv\{z/N:z\in\mathcal Z_N(b)\}.
\]
Thus $P_N(b)\subseteq P(b)$. Both are compact polytopes when nonempty, since their coordinates are nonnegative and sum to one. The assumption that $(N,B)$ is feasible implies that $\mathcal Z_N(b)$, and therefore $P_N(b)$, are nonempty. 

Given $\pi\in\Pi_{N,B}$, its induced mass vector is
\[
y^\pi_{\theta s} =\frac1N\sum_{i:\theta_i=\theta}\Pr_\pi(\tilde s_i=s).
\]
Every deterministic realization has a count table in $\mathcal Z_N(b)$, so $y^\pi\in P_N(b)$. If $\pi$ treats identical types equally, then $y^\pi_{\theta s}=d_\theta x_{\theta s}$.

For a mass vector $y\in P(b)$ and a type with $d_\theta>0$, let $x_\theta=\frac{y_\theta}{d_\theta}\in\Delta(\bun)$.  The scalar $u_\theta\cdot x_\theta$ is the expected utility of an agent of type $\theta$. The entries of $y_\theta$ are population masses, and $u_\theta\cdot y_\theta=d_\theta(u_\theta\cdot x_\theta)$ is that type's contribution to population-average utilitarian welfare.

\begin{table}[tt]
\begin{center}
\small
\renewcommand{\arraystretch}{1.2}
\begin{tabular}{@{}p{0.24\linewidth}p{0.72\linewidth}@{}}
\hline
\textbf{Notation} & \textbf{Meaning} \\
\hline
$\pi$ & A joint lottery over feasible deterministic assignments. \\
$x_i$ & Agent $i$'s marginal lottery over bundles under $\pi$. \\
$x_\theta$ & The common marginal lottery for type $\theta$, when agents
of that type are treated equally. \\
$z\in\mathcal Z_N(b)$ & An integer count table for a deterministic
assignment with $z/N$ being a ``normalized'' count table. \\
$y\in P_N(b)$ & An implementable mass vector: expected normalized
counts under a random feasible allocation. \\
$y\in P(b)$ & A mass vector satisfying the type and resource
constraints. \\
\hline
\end{tabular}
\end{center}
\caption{Summary of notation for allocations} \label{tab:notation}
\end{table}
 
Table~\ref{tab:notation} summarizes the notation we have introduced. Note the tension between these different objects. A mass vector $y\in P(b)$ corresponds to allocations in a large economy: for any $y\in P(b)$, type $\theta\in \Theta$ and bundle $s\in \bun$ there exists a deterministic assignment in the large economy in which a mass $y_{\theta s}$ agents of type $\theta$ receive bundle $s$. In contrast, the elements of $P_N(b)$ are fractional vectors that can be traced back to random feasible allocations in a finite economy with $N$ agents.

\begin{samepage}
\begin{lemma}\label{lem:implementation}
Let $y\in P(b)$. There exists a random feasible allocation $\pi\in \Pi_{N,B}$ satisfying
\[
\Pr_\pi(\tilde s_i = s) = \frac{y_{\theta_i s}}{d_{\theta_i}} \qquad(i=1,\ldots,N,\ s\in\mathcal S)
\]
if and only if $y\in P_N(b)$.
\end{lemma}
\end{samepage}

\begin{proof}
Every deterministic allocation produces a count table in $\mathcal Z_N(b)$; taking expectations proves necessity. Conversely, write $y=\sum_\ell\lambda_\ell z^\ell/N$, where $z^\ell\in\mathcal Z_N(b)$. First draw $\ell$ with probability $\lambda_\ell$. For each present type $\theta$, form a list containing $z^\ell_{\theta s}$ copies of bundle $s$ and assign this list by a uniform random permutation to that type's $n_\theta$ agents. Every realization is feasible, and each agent of that type receives $s$ with probability
\[
\sum_\ell\lambda_\ell\frac{z^\ell_{\theta s}}{n_\theta} = \frac{y_{\theta s}}{d_\theta}=x_{\theta s}.
\]
\end{proof}
The feasible mass vectors in a large economy are the elements of $P(b)$. In a finite economy, $P(b)$ is a relaxation of $P_N(b)$ (the mass vectors that are implementable); in a large economy, $P(b)$ is the allocation set itself.  Now we argue that, in many of the applications we are concerned with, we may have $P_N(b)\neq P(b)$.

\subsection{Applications}\label{sec:applications}

We have emphasized that our results apply to settings where one cannot work with marginal assignments that satisfy the feasibility constraints in expectation. We proceed to describe some of the applications that are special cases of our model, but where the bihierarchy condition of \citet{BudishEtAl2013} fails:  A hierarchy is a family of sets whose members are disjoint or nested; a bihierarchy is a family that can be partitioned into two hierarchies. Three pairwise intersecting sets, none containing another, cannot belong to a bihierarchy. The condition is sufficient in general and necessary for implementability in ``canonical'' two-sided assignment problems. For general constraint structures, failure of bihierarchy alone does not imply failure of implementability. The combinatorial assignment example below gives an explicit failure.

Despite the emphasis on settings where marginal assignments are problematic, our results do apply to the Hylland-Zeckhauser model where the bihierarchy condition is satisfied. We work this out in Section~\ref{sec:HZ}, which also shows that we obtain a quantitative improvement over Theorem~\ref{thm:main} for this case.

\subsubsection{Multi-period allocations}\label{ex:birkhoff-failure}

Consider a finite set  $O$ of objects and a finite number of periods $T$. Agents consume bundles that consist of a sequence of objects, $(o_1,\dots,o_T)$. Let $\bun=O^T$. At date $t$, there are $B_{to}$ physical copies of object $o$, with $\sum_oB_{to}=N$. Define $R_{(t,o),s}=\one\{s_t=o\}$. Then feasibility requires that, in every period, each agent receives one object and that every available object is assigned. Specifically, if agent $i$ consumes a sequence of objects $(o^1_i,\dots,o^T_i)\in \bun$ then we require that $\sum_{i=1}^N \one_{o^t_i=o'} = B_{to'}$ for $t=1,\dots,T$ and all $o'\in O$. 

To see that implementability can fail, consider a simple example with $N=T=2$ and $O=\{a,b\}$. Let $\bun=O^T$. Each agent has a
different type, labeled $1$ and $2$. At each date, there is one copy of each object. Resource uses are $R_{(t,o),s}=\one\{s_t=o\}$, and $B_{to}=1$ for every date $t$ and object $o$. Consider the marginal distributions
\[
x_1(a,a)=x_1(b,b)=\frac12,\qquad x_2(a,b)=x_2(b,a)=\frac12,
\]
with all other probabilities equal to zero.

These distributions satisfy the period-by-period resource constraints in expectation. In each period, each agent receives each object with probability $1/2$, so the fractional assignment matrix is
\[
\begin{pmatrix}1/2&1/2\\1/2&1/2\end{pmatrix}.
\]
Since $d_1=d_2=1/2$, the associated mass vector $y_1=x_1/2$,
$y_2=x_2/2$ belongs to $P(b)$. Nevertheless, these marginals cannot be induced by a random feasible allocation.

There are four feasible deterministic allocations:
\[
\begin{array}{c|c}
\text{Agent 1's sequence}&\text{Agent 2's sequence}\\\hline
(a,a)&(b,b)\\
(a,b)&(b,a)\\
(b,a)&(a,b)\\
(b,b)&(a,a)
\end{array}
\]
Because $x_1(a,b)=x_1(b,a)=0$, any lottery inducing $x_1$ must put probability zero on the second and third allocations. Under each of the remaining allocations, agent $2$ receives the same object in both periods. Its marginal distribution must therefore be supported on $\{(a,a),(b,b)\}$, contrary to the definition of $x_2$.

Thus $y\in P(b)$ but $y\notin P_N(b)$. The missing restrictions concern the joint implementability of the agents' marginal distributions.

\subsubsection{Course allocation with timetable and curricular constraints}

Let $C$ be a finite set of courses, and let $\bun\subseteq2^C$ be the common set of admissible course schedules. For each course $c$, set $R_{cs}=\one\{c\in s\}$, and let $B_c$ be its supply of seats. As our feasibility constraints are formulated as equalities, we consider the case in which all available seats are assigned. Common timetable and curricular restrictions can be incorporated directly into $\bun$. \citet[Sections~II.A and~IV.A]{BudishEtAl2013} discuss course allocation and explain the difficulty of combining these two kinds of
restrictions.

Suppose there are finance courses $f_1,f_2$ and marketing courses $m_1,m_2$, where the subscript indicates a time slot. Require every student to take at most one course in each time slot and at most one course in each subject. In their agent--course coordinates, the constraint sets include
\[
\{i\}\times\{f_1,m_1\},\qquad \{i\}\times\{f_1,f_2\},\qquad \{1,\ldots,N\}\times\{f_1\}.
\]
The first two sets describe a timetable constraint and a curricular constraint for student $i$; the third describes the capacity of course $f_1$. When $N\geq2$, these sets are pairwise intersecting and nonnested, so the constraint structure is not a bihierarchy. Our formulation accommodates the common restrictions by treating each admissible schedule as a bundle.

\subsubsection{Joint assignment of different goods}

An agent may receive a combination of goods or services: housing and parking, an office and equipment, a lecture and a laboratory section, or a school and an after-school program. Let $H$ and $L$ be two disjoint finite sets of object categories, and take $\bun=H\times L$. A bundle $(h,\ell)$ uses one unit of category $h$ and one unit of category $\ell$. Utilities $u_\theta(h,\ell)$ depend on the combination of the good and service consumed.

In agent--bundle coordinates, the constraint sets include those for each agent, each category in $H$, and each category in $L$. This is the  three-sided assignment structure studied by \citet[Section~VI.A, Theorem~12]{BudishEtAl2013}. With at least two agents and two categories in each of $H$ and $L$, an agent constraint, an $H$-category constraint, and an $L$-category constraint are pairwise intersecting and nonnested. Thus the structure is not a bihierarchy. Each complete combination is nevertheless a bundle in our model, with resource uses independent of the recipient.

\subsubsection{General combinatorial assignment}

Let $G$ be a finite set of goods or resource components, and let $\bun\subseteq2^G$ be a common collection of admissible bundles.
For $h\in G$, define
\[ R_{hs}=\one\{h\in s\}.
\]
The supply $B_h$ is the number of available copies of good $h$. Examples include the assignment of equipment kits and packages of computing resources. Course allocation is a special case: the goods are courses, and $\bun$ consists of the admissible schedules.

Overlapping resource requirements can violate the bihierarchy condition. For example, let $G=\{a,b,c\}$ and $\bun=2^G$. In the count-table
coordinates, the resource constraint sets are
\[
H_h=\{(\theta,s)\in\Theta\times\bun:h\in s\}, \qquad h\in\{a,b,c\}.
\]
The sets $H_a,H_b,H_c$ are pairwise intersecting and nonnested. The bundles $\{a,b\}$, $\{b,c\}$, and $\{a,c\}$ witness the three intersections, each belonging to exactly two of the resource sets. This also gives an odd cycle in the sense of \citet[Section~VI, Definition~4]{BudishEtAl2013}.

There can be an actual failure of implementation. Take $N=2$, with one agent of type $1$ and one of type $2$, and supplies $B=(1,1,1)$. Consider the marginal lotteries
\[
x_1=x_2= \tfrac14\delta_{\varnothing} +\tfrac14\delta_{\{a,b\}} +\tfrac14\delta_{\{b,c\}} +\tfrac14\delta_{\{a,c\}},
\]
where $\delta_s$ denotes the point mass at bundle $s$. Each agent receives each good with probability $1/2$, so aggregate resource use equals $B$ in expectation. Hence, the associated mass vector $y_\theta=x_\theta/2$ belongs to $P(b)$, with $d=(1/2,1/2)$ and $\sigma=B/2$.

In every feasible realization, at most one agent can receive a two-good bundle because any two such bundles share a good. The proposed marginals instead imply an expected number of two-good bundles equal to $2(3/4)=3/2$. Thus, they cannot be implemented, and $y\notin P_N(b)$. The economy is feasible: for every $s\subseteq G$, the allocation $(s,G\setminus s)$ assigns all goods exactly once. The same obstruction applies if resources may remain unused.

The object to be implemented is the distribution over complete bundles. Implementability of the induced item marginals alone does not ensure the implementation of these distributions, which are needed to evaluate arbitrary bundle utilities $u_\theta(s)$.

\section{Efficiency, fairness and Lindahl equilibrium}\label{sec:existence}

We present a common existence result for both finite and large economies. This has two advantages. First, it makes clear that the existence argument is not specific to either finite or large economies. Second, it provides the large-market result we need to establish approximate incentive compatibility in large finite economies.

Throughout this section, we assume $d\gg0$. If $d_\theta=0$ for some type $\theta$, we can simply omit that type from the existence results. For this unified treatment, we extend the finite-economy notions of Pareto optimality, envy-freeness, and Lindahl equilibrium from Section~\ref{sec:model} to large economies.

\begin{samepage}
The notions of Pareto optimality and envy-freeness extend to $P(b)$  using the expected utilities of each type. A mass vector $y\in P(b)$ is \df{Pareto optimal in $P(b)$} if there is no $y'\in P(b)$ such that
\[
u_\theta\cdot\frac{y'_\theta}{d_\theta} \geq u_\theta\cdot\frac{y_\theta}{d_\theta} \quad(\theta\in\Theta),
\]
with strict inequality for at least one type.  
It is \df{$\epsilon$-envy-free} if $u_\theta\cdot\frac{y_\theta}{d_\theta}+\epsilon \geq u_\theta\cdot\frac{y_\tau}{d_\tau}$ for all $\theta,\tau\in\Theta$.
\end{samepage}

We similarly express Lindahl equilibrium in terms of mass vectors. Prices and incomes in this definition are per agent of each type, so aggregate revenue weights them by the corresponding population masses.

\begin{samepage}
\begin{definition}\label{def:lindahl-mass}
A \df{Lindahl equilibrium on $P(b)$} is a tuple $(y,p,w)$, where
$y\in P(b)$, $p=(p_\theta)_{\theta\in\Theta}$ with
$p_\theta\in\Re^{qm}$, and $w\in\Re_+^q$, such that:
\begin{enumerate}
    \item For every type $\theta$,
    \[
    y\in\argmax\left\{
    u_\theta\cdot\frac{y'_\theta}{d_\theta}:
    p_\theta\cdot y'\leq w_\theta,\ y'\in P(b)\right\},
    \]
    and $y$ minimizes $p_\theta\cdot y'$ among all solutions to this
    maximization problem.
    \item
    \[
    y\in\argmax_{y'\in P(b)}
    \left(\sum_{\theta\in\Theta}d_\theta p_\theta\right)\cdot y'.
    \]
\end{enumerate}
\end{definition}
\end{samepage}

\begin{theorem}\label{thm:existence_lindahl}\label{thm:prelim} 
Let $b=(d,\sigma)$ with $d\gg0$.
Let $Q=P(b)$, with $P(b)$ nonempty, or let $Q=P_N(b)$ for a feasible finite economy with type distribution $d$ and supply vector $B=N\sigma$.

There exist a mass vector $y^*\in Q$ and strictly positive numbers $a_\theta$ such that the associated marginal assignments $x_\theta^*=y_\theta^*/d_\theta$ are envy-free:
\[
u_\theta\cdot x_\theta^*\geq u_\theta\cdot x_\tau^*
\qquad(\theta,\tau\in\Theta).
\]
Moreover, $y^*$ solves
\begin{equation}\label{eq:weighted-welfare}
\max_{y\in Q}\sum_{\theta\in\Theta}
d_\theta a_\theta\bigl(u_\theta\cdot x_\theta\bigr),
\qquad x_\theta=\frac{y_\theta}{d_\theta}.
\end{equation}

When $Q=P(b)$, $y^*$ is Pareto optimal in the large economy, and every Pareto optimal mass vector in $P(b)$ can be supported as a Lindahl equilibrium in the sense of Definition~\ref{def:lindahl-mass}.

When $Q=P_N(b)$, $y^*$ can be implemented by a random feasible allocation $\pi^*\in\Pi_{N,B}$ that treats identical types equally, is Pareto optimal and envy-free, and can be supported as a Lindahl equilibrium in the sense of Definition~\ref{def:lindahl-finite}. Every Pareto optimal $\pi\in\Pi_{N,B}$ can also be supported as such an equilibrium.
\end{theorem}

Our large-market incentive compatibility result constructs a mechanism using the allocation $y^*$ and weights $a_\theta$ supplied by Theorem~\ref{thm:prelim} when applied to the reference economy $b^*$, with $Q=P(b^*)$.

\section{Approximate incentive compatibility}\label{sec:mainSPthm}

\subsection{Mechanisms}

Consider a pair $(N,B)$ that is feasible. For each type profile $(\theta_{i})_{i=1}^N\in\Theta^N$, we have a finite economy $((\theta_{i})_{i=1}^N,B)$. An agent's type is private information, but the utility $u_{\theta_i}$ defined by their type is known once $\theta_i$ is determined. We are interested in the allocations that can be obtained given the presence of private information. 

A (direct revelation)  \df{mechanism} is a function $m \colon \Theta^N \to \Pi_{N,B}$  that  assigns a lottery over feasible deterministic allocations to every report profile $r\in \Theta^N$. Given a mechanism $m$ and a report $r\in\Theta^N$, let $m(r)_i$ be the marginal distribution over $\bun$ corresponding to agent $i$ (this would be the distribution of the random variable $\tilde s_i$). A mechanism \df{treats equal types equally}, or is \df{anonymous}, if any two agents that report the same type obtain the same lottery on $\bun$. For an anonymous mechanism $m$, we shall write $m(r)_{\theta}$ to denote the marginal distribution over $\bun$ obtained by any agent that reports type $\theta$.

For $\epsilon \geq 0$, a mechanism $m$ is \df{$\epsilon$-incentive compatible at the type profile $\theta=(\theta_i)_{i=1}^N$} if, for every agent $i$ and every alternative report $\tau\in\Theta$, writing $\pi=m(\theta)$ and $\pi'=m(\tau,\theta_{-i})$, we have
\[
 \mathbb E_{\pi}[u_{\theta_i}(\tilde s_i)] +\epsilon \geq  \mathbb E_{\pi'}[u_{\theta_i}(\tilde s_i)].
\]

The mechanism is \df{$\epsilon$-incentive compatible} if this condition holds at every type profile $\theta\in\Theta^N$. 

For $\epsilon \geq 0$, a mechanism $m$ is \df{$\epsilon$-envy-free at the type profile $\theta=(\theta_i)_{i=1}^N$} if the random allocation $m(\theta)$ is $\epsilon$-envy-free. It is \df{$\epsilon$-envy-free} if $m(\theta)$ is $\epsilon$-envy-free for all $\theta\in\Theta^N$.

A mechanism $m$ is \df{Pareto optimal} if, for every type profile $\theta\in \Theta^N$, the random allocation $m(\theta)$ is Pareto optimal.

\subsection{Definition of the mechanism}\label{sec:mechanism}

We begin by fixing a large-economy limit and then consider finite economies whose type distributions and per-agent supplies are close to those in the limit economy. The large economy provides a benchmark solution and a welfare criterion. In a finite economy, the mechanism selects among the implementable mass vectors that maximize the welfare criterion, one which is closest to the benchmark. The approximation of finite economies to the large-economy limit is discussed briefly in Section~\ref{sec:asymptotics}; standard replica and sampling arguments provide natural ways of generating such approximations. Here we describe the mechanism and state the approximate incentive compatibility result.

To define our mechanism, we start from an allocation in the limiting large economy. The allocation has the desired normative properties of Pareto optimality and lack of envy. To this end, fix a large economy $b^*=(d^*,\sigma^*)$ with $d^*\gg0$ and $P(b^*)\neq\varnothing$. Apply Theorem~\ref{thm:existence_lindahl} with $b=b^*$ and $Q=P(b^*)$. Choose once and for all a mass vector $y^*$ and strictly positive weights $a=(a_\theta)_{\theta\in\Theta}$ supplied by that theorem. The marginal lotteries $x_\theta^*=y_\theta^*/d_\theta^*$ are envy-free, and
\[
y^*\in\argmax_{y\in P(b^*)} \sum_{\theta\in\Theta} a_\theta u_\theta \cdot y_{\theta}.\footnote{Recall that $u_\theta\cdot y_\theta = \sum_s u_\theta(s)y_{\theta s}$.}
\]
This is the objective in the existence theorem, since $d_\theta^*a_\theta(u_\theta\cdot x_\theta) = a_\theta u_\theta\cdot y_\theta$.

The pair $(y^*,a)$ remains fixed across population sizes, supplies, and report profiles. Set
\[
F(b)=\argmax_{y\in P(b)} \sum_{\theta\in \Theta} a_\theta u_\theta \cdot y_\theta
\]
whenever $P(b)$ is nonempty. In particular, $y^*\in F(b^*)$. Recall that $P(b)$ is defined entirely in terms of the normalized economy $b$ and does not depend on any particular finite realization $(N,B)$.

For each feasible pair $(N,B)$, we define a mechanism $m_{N,B}:\Theta^N\to\Pi_{N,B}$. Given a report profile $r$, let
\[
n_\theta(r)=|\{i:r_i=\theta\}|,\qquad d_\theta(r)=\frac{n_\theta(r)}N,\qquad b(r)=(d(r),B/N).
\]
The idea is to approximate $y^*$ among the implementable mass vectors that maximize the fixed welfare objective in the finite economy. For $b=b(r)$, let
\begin{align}
\mathcal Z_N^{\rm opt}(b) &=\argmax_{z\in\mathcal Z_N(b)} \sum_{\theta\in \Theta }a_\theta u_\theta \cdot z_\theta, \label{eq:zopt}\\
F_N(b) &=\conv\{z/N:z\in\mathcal Z_N^{\rm opt}(b)\}, \label{eq:finite-face}\\
y^N(b) &=\argmin_{y\in F_N(b)}\norm{y-y^*}^2. \label{eq:mechanism}
\end{align}
The set $\mathcal Z_N(b)$ is nonempty at every report profile: take a feasible deterministic allocation and record the bundles received by agents of each reported type. The set $F_N(b)$ is precisely the set of mass vectors in $P_N(b)$ that maximize the fixed welfare objective. Thus $F_N(b)$ is a nonempty compact convex set, and the last minimizer (a projection onto $F_N(b)$) is unique.

Since $y^N(b)\in F_N(b)\subseteq P_N(b)$, Lemma~\ref{lem:implementation}, applied to reported types, implements it as a random feasible assignment that gives agents who report the same type the same marginal assignments. Let $m_{N,B}(r)$ be such a random assignment. For every reported type $\theta$, write $x_\theta^N(b)=\frac{y_\theta^N(b)}{d_\theta(r)}$.

\begin{theorem}\label{thm:main}
For every feasible pair $(N,B)$, the mechanism $m_{N,B}$ treats equal types equally and is Pareto optimal.

For every $\epsilon>0$, there exist $\eta>0$ and $N_\epsilon\in\mathbb N$ such that, for every feasible pair $(N,B)$ with $N\geq N_\epsilon$, the mechanism $m_{N,B}$ is $\epsilon$-envy-free and $\epsilon$-incentive compatible at every type profile $(\theta_1,\dots,\theta_N)\in\Theta^N$ satisfying
\[
 \bigl\|(d(\theta_1,\dots,\theta_N),B/N)-b^*\bigr\|\leq\eta.
\]
\end{theorem}

The definition of the mechanism $m_{N,B}$ depends on a given large economy $b^*=(d^*,\sigma^*)$, and hence on a type distribution $d^*$. When $b^*$ represents the limiting or population economy, this requires knowledge of the population distribution $d^*$. This is analogous to Bayesian incentive compatibility, but the incentive compatibility obtained is stronger. There is no assumption of a common prior or equilibrium behavior, and payoffs are not averaged across types (so incentive compatibility is ``pointwise.'').

Fix an agent $i$ with true type $\theta_i$ and arbitrary reports $r_{-i}$ by the other agents such that $(d(\theta_i,r_{-i}),B/N)$ is sufficiently close to $b^*$. If the population is sufficiently large, then truthful reporting is an $\epsilon$-best response to $r_{-i}$. The other agents may misreport their types: the restriction is on the empirical distribution of their reports, rather than on whether those reports are truthful or form an equilibrium.

\section{Discussion}\label{sec:discussion}

\subsection{Large but finite economies}\label{sec:asymptotics}

The message of Theorem~\ref{thm:main} for approximate incentive compatibility requires that a finite economy with many agents approximates the large-economy limit, in the sense that we have $\|(d(\theta_1,\dots,\theta_N),B/N)-b^*\|\leq\eta$. This can be justified using standard ideas.

First, a common way to model economies with many agents is through the idea of replicas. If $d^*$ is rational, for example, then we can let the $k$-fold replica of an economy with $n^0(\theta)$ agents of type $\theta$ and $N^0=\sum_\theta n^0(\theta)$ be such that $N=kN^0$ and $n_\theta=kn_\theta^0$. So we get $d_\theta=\frac{n_\theta^0}{N^0}=d_\theta^*$

Second, we may consider a finite economy to be an iid sample from the limiting economy with type distribution $d^*$. Then the sample (empirical) distribution of types satisfies
\[
d_\theta^{(N)} =\frac1N\sum_{i=1}^N\one\{\theta_i=\theta\},
\] so we obtain $d_\theta^{(N)}\to d^*_\theta$ a.s.\ as a consequence of the strong law of large numbers. It is also easy to show that 
\[
\Pr\!\left(\|d^{(N)}-d^*\|>\eta\right) \leq\frac{1}{N\eta^2} 
\] using standard concentration inequalities.

\subsection{Unit-demand assignment}\label{sec:HZ}

As an illustration, we show how our model specializes to the unit-demand assignment model of \citet{HyllandZeckhauser1979}. The set $\bun$ now consists of object categories, and each agent receives exactly one object. The resources are identified with the same set of categories, so we can index them by the set of object categories and let $R_{hs}=\one\{h=s\}$. There are $B_s$ copies of object $s$.  Feasibility requires that 
$\sum_{s\in\bun}B_s=N$. Thus
\[
P(b) = \left\{ y\geq0: \sum_{s\in\bun}y_{\theta s}=d_\theta \quad(\theta\in\Theta),\qquad \sum_{\theta\in\Theta}y_{\theta s}=\sigma_s \quad(s\in\bun)
\right\},
\]
where $Nd_\theta=n_\theta$ and $N\sigma_s=B_s$ are integers. The sets of object categories and utility types remain fixed as the population and the number of copies grow. 

The Birkhoff--von Neumann theorem implies that $P_N(b)=P(b)$, in contrast to the examples in Section~\ref{sec:applications}. To see this, take $y\in P(b)$ and give every agent of type $\theta$ the marginal lottery $x_\theta=y_\theta/d_\theta$. Split each object category $s$ with $B_s>0$ into its $B_s$ physical copies, assigning a probability $x_{\theta s}/B_s$ to each copy for an agent of type $\theta$. Categories with zero supply can be omitted, since their columns in $y$ are zero.

The resulting matrix of probabilities has one row per agent and one column per physical copy. It is an $N\times N$ doubly stochastic matrix: each row sums to one, and each column corresponding to a copy of $s$ sums to
\[
\frac1{B_s} \sum_{\theta:d_\theta>0}n_\theta x_{\theta s} = \frac{N}{B_s}\sum_{\theta\in\Theta}y_{\theta s} =1.
\]
By Birkhoff--von Neumann, these probabilities can be implemented through a lottery over permutations assigning every physical copy to exactly one agent. Every realization induces a count table $z\in\mathcal Z_N(b)$, and its expected counts satisfy
\[
\mathbb E[z_{\theta s}] =n_\theta x_{\theta s} =Ny_{\theta s}.
\]
Hence $y=\mathbb E[z/N]\in P_N(b)$. The reverse inclusion always holds, proving the equality.

It follows that the fractional and implementable welfare-optimal sets also coincide:
\[
F_N(b)=F(b) =\argmax_{y\in P(b)} \sum_{\theta\in\Theta}\sum_{s\in\bun} a_\theta u_\theta(s)y_{\theta s}.
\]
Indeed, a convex decomposition of a welfare maximizer into normalized integer count tables can assign positive probability only to welfare-maximizing count tables. Consequently, the mechanism constructed above selects
\[
y^N(b)=\argmin_{y\in F(b)}\norm{y-y^*}^2.
\]
The sensitivity bound in Lemma~\ref{lem:optimal-sets} therefore
gives
\[
\norm{y^N(b)-y^*} =\dist(y^*,F(b)) \leq H\norm{b-b^*},
\]
with the constant $H$ supplied by that lemma for the fixed welfare objective. (The integer-proximity step in our proof is unnecessary.)

Theorem~\ref{thm:main} applies to this specialization of the Hylland--Zeckhauser environment. For every feasible finite economy, the mechanism treats identical reports equally and selects a feasible lottery that is Pareto optimal with respect to reported utilities. For every $\epsilon>0$, there exist $\eta>0$ and $N_\epsilon\in\mathbb N$ such that the mechanism is $\epsilon$-envy-free and $\epsilon$-incentive compatible at every type profile satisfying
\[
N\geq N_\epsilon, \qquad \norm{(d(\theta_1,\ldots,\theta_N),B/N)-b^*}\leq\eta.
\]

The bounds in this case are sharper because there is no discrepancy between $P_N(b)$ and $P(b)$. In particular, whenever a finite economy has
$b=b^*$, the mechanism selects $y^N(b)=y^*$ and is exactly envy-free. A unilateral report changes the empirical type distribution by at most $\sqrt2/N$, so the sensitivity and normalization arguments in the proof of Theorem~\ref{thm:main} bound any gain from manipulation by $O(N^{-1})$. More generally, uniformly for sufficiently large economies near $b^*$, envy is bounded by $O(\norm{b-b^*})$ and gains from manipulation by $O(\norm{b-b^*}+N^{-1})$.

\subsection{Scope of the feasibility representation}\label{sec:feasibility}

Our representation of the feasible allocations requires spelling out the resource use of each bundle. As we have seen in Section~\ref{sec:applications}, this covers a lot of ground. It is also clear that it imposes limits; for example, resource use is bundle-specific and cannot depend on who the agent receiving the bundle is. It rules out that giving bundle $s$ to one agent may require more resources than giving it to a different agent. Here we discuss the scope of our representation. What is and what is not allowed in our model?

Fix a population size $N\geq1$ and a menu $\bun$ of $m\geq1$ bundles. An aggregate bundle-count vector records the number $t_s$ of agents receiving each bundle $s$. The possible count vectors form $H_N=\{t\in\Z_+^m \colon \one \cdot t = N\}$. The next result is a simple observation stating that $\mathcal C$ comes from a feasible set of allocations only when it is closed under affine transformations that remain in $H_N$: so, to the extent that an affine transformation is compatible with the number of agents and integer constraints inherent in the indivisibility of the bundles, it should arise from a feasible allocation.

\begin{proposition}\label{prop:resource-representation}
Let $\mathcal C\subseteq H_N$ be nonempty. There exist an integer $k\geq1$, a matrix $R\in\Z_+^{k\times m}$, and a vector $B\in\Z_+^k$ such that $\mathcal C=\{t\in H_N:Rt=B\}$ if and only if $\mathcal C=\operatorname{aff}(\mathcal C)\cap H_N$.

Moreover, these resource equations can be chosen so that, for every $t\in\Re^m$ satisfying $\one \cdot t=N$, the equality $Rt=B$ holds if and only if $t\in\operatorname{aff}(\mathcal C)$.
\end{proposition}

\begin{proof}
Suppose first that $\mathcal C=\{t\in H_N:Rt=B\}$. Every equation in $Rt=B$ holds throughout $\operatorname{aff}(\mathcal C)$, because it holds at every point of $\mathcal C$. Thus $\operatorname{aff}(\mathcal C)\cap H_N\subseteq\mathcal C$. The reverse inclusion follows from $\mathcal C\subseteq H_N$.

Conversely, suppose that $\mathcal C=\operatorname{aff}(\mathcal C)\cap H_N$, and choose $t^0\in\mathcal C$. Consider the homogeneous system $g^\top(t-t^0)=0$ for all $t\in\mathcal C$, with unknown $g\in\Re^m$. This is a finite system with integer coefficients. Gaussian elimination gives a rational basis of its solution space. Taking common denominators in each basis vector gives a basis consisting of integer vectors. 

Let $G\in\Z^{k\times m}$ have these basis vectors as its rows. There is at least one row, since $\one$ solves the system. The kernel of $G$ is the span of the vectors $t-t^0$, $t\in\mathcal C$. Consequently, $Gt=Gt^0$ holds exactly when $t\in\operatorname{aff}(\mathcal C)$.

Choose a nonnegative integer $\ell$ sufficiently large that $R_{js}=G_{js}+\ell\geq0$ for every row $j$ and bundle $s$, and set $B=Rt^0$. Both $R$ and $B$ have nonnegative integer entries. For every $t\in\Re^m$ satisfying $\one \cdot t=N$, we have $Rt-B=G(t-t^0)$: the constant added to each row contributes $\ell \one \cdot (t-t^0)=0$. Hence $Rt=B$ holds exactly when $t\in\operatorname{aff}(\mathcal C)$ on this hyperplane. This proves the additional assertion. Restricting to $H_N$ and using the hypothesis gives $\mathcal C=\{t\in H_N:Rt=B\}$.
\end{proof}

The proposition also characterizes which sets of deterministic
allocations admit our resource representation.

\begin{corollary}\label{cor:allocation-representation}
Let $\mathcal F\subseteq\bun^N$ be nonempty, and let
$\mathcal C\subseteq H_N$ consist of its aggregate bundle-count
vectors. There exist common resource requirements
$R\in\Z_+^{k\times m}$ and supplies $B\in\Z_+^k$ such that
an allocation $(s_1,\ldots,s_N)$ belongs to $\mathcal F$ exactly
when $\sum_iR_{s_i}=B$ if and only if $\mathcal F$ is invariant
under permutations of agents and
$\mathcal C=\operatorname{aff}(\mathcal C)\cap H_N$.
\end{corollary}

\begin{proof}
A common resource representation is invariant under permutations
of agents, since these leave total resource use unchanged.
Its feasible count vectors are exactly
$\{t\in H_N:Rt=B\}$, so
Proposition~\ref{prop:resource-representation} gives the
affine-hull condition.

Conversely, suppose that both conditions hold. By the proposition,
choose $R$ and $B$ such that
$\mathcal C=\{t\in H_N:Rt=B\}$. Any two allocations with the
same aggregate bundle counts are permutations of one another.
Permutation invariance therefore implies that an allocation
belongs to $\mathcal F$ exactly when its count vector $t$
belongs to $\mathcal C$. Since $\sum_iR_{s_i}=Rt$, this is
equivalent to $\sum_iR_{s_i}=B$.
\end{proof}

Proposition~\ref{prop:resource-representation} concerns a fixed $N$. A different question is to characterize feasibility across population sizes. We have not pursued this here. We should also emphasize that we can work with subsets of the set of feasible allocations, as long as they are permutation invariant.

\section{Proofs}\label{sec:proof}

\subsection{Proof of Theorem~\ref{thm:existence_lindahl}}
Throughout this subsection, $Q$ denotes either $P(b)$ or $P_N(b)$, as in Theorem~\ref{thm:existence_lindahl}.

\begin{lemma}\label{lem:cycle}
  A \df{cycle} of types is a sequence $(\theta_1,\ldots,\theta_L)$ of distinct types, with $L\geq2$. We regard its indices cyclically, so $\theta_{L+1}=\theta_1$. Let $y\in Q$, and $(\theta_1,\ldots,\theta_L)$ be a cycle. Write $x_\theta=y_\theta/d_\theta$. There exists $t_0>0$ such that, for every $0\leq t\leq t_0$, the mass vector obtained by replacing
\[
y_{\theta_j}\quad\hbox{with}\quad y_{\theta_j}+t(x_{\theta_{j+1}}-x_{\theta_j}) \qquad(j=1,\ldots,L)
\]
and leaving all other rows unchanged belongs to $Q$. For $Q=P(b)$ one may take $t_0=\min_jd_{\theta_j}$, and for $Q=P_N(b)$ one may take $t_0=1/N$.
\end{lemma}

\begin{proof}
Suppose first that $Q=P(b)$. For $0\leq t\leq\min_jd_{\theta_j}$, each modified row can be written as
\[
(d_{\theta_j}-t)x_{\theta_j}+t x_{\theta_{j+1}}.
\]
Since each $x_\theta$ is a probability vector over $\mathcal S$, this row is nonnegative and has total mass $d_{\theta_j}$.
The sum of the changes across the cycle is zero, so aggregate bundle masses, and therefore resource use, are unchanged. Hence the modified vector satisfies \eqref{eqP}, and therefore belongs to $P(b)$.

Now suppose that $Q=P_N(b)$. By Lemma~\ref{lem:implementation}, choose $\pi\in\Pi_{N,B}$ giving every type-$\theta$ agent the marginal lottery $x_\theta$. Independently of the allocation, select one agent uniformly from each type in the cycle. In every realization, give the selected type-$\theta_j$ agent the bundle received by the selected type-$\theta_{j+1}$ agent, and leave other agents' bundles unchanged. This operation preserves the collection of assigned bundles and hence feasibility.

The selected type-$\theta_j$ agent's original bundle has marginal $x_{\theta_j}$, and its replacement has marginal $x_{\theta_{j+1}}$. The new random assignment $\pi'$ has mass vector $y'$, whose type-$\theta_j$ row is $y_{\theta_j}+\frac1N(x_{\theta_{j+1}}-x_{\theta_j})$. For $0 \leq t \leq 1/N$ define the distribution $\pi^t = (1-Nt) \pi + Nt\,\pi'$ that mixes $\pi'$ with probability $Nt$ and $\pi$ with probability $1-Nt$. Every realization of this randomization over deterministic allocation is feasible, because this holds for $\pi$ and $\pi'$. For each type in the cycle,  
\[
\begin{aligned}
y^{\pi^t}_{\theta_j}
&=(1-Nt)y_{\theta_j}+Nt\,y'_{\theta_j}\\
&=(1-Nt)y_{\theta_j}
  +Nt\left[y_{\theta_j}
  +\frac1N(x_{\theta_{j+1}}-x_{\theta_j})\right]\\
&=y_{\theta_j}+t(x_{\theta_{j+1}}-x_{\theta_j}).
\end{aligned}
\]
All other rows remain unchanged. Hence the desired mass vector is induced by a feasible lottery and belongs to \(P_N(b)\). 
\end{proof}

Our next result is a version of the polyhedral support result in \cite{McLennan2002}. The existence of strictly positive weights is as in \cite{Carroll2010}. See also \cite{echenique2022efficiencyrandomresourceallocation} for an overview of these results

\begin{lemma}\label{lem:positive-support}
Let $C$ be a nonempty compact polytope and let $U=(U_1,\ldots,U_r):C\to\mathbb R^r$ be a profile of linear utilities. Every Pareto optimal $z\in C$ maximizes a strictly positive weighted sum of these utilities. The set of Pareto optimal points of $C$ is closed.
\end{lemma}

\begin{proof}
Let $z$ be Pareto optimal, write $V=U(C)$ and $v=U(z)$, and consider $K=\cone(V-v)$. Since $C$ is a polytope and $U$ is linear, $K$ is a finitely generated polyhedral cone, and hence closed. Pareto optimality implies $K\cap\Re_+^r=\{0\}$: otherwise a positive rescaling of a nonzero nonnegative member of $K$ would give a Pareto improvement in $V$.

The closed cone $K$ and the compact simplex $\Delta^{r-1}$ are disjoint and can be strictly separated. Thus there exists $a\in\Re^r$ with
\[
\sup_{h\in K}a\cdot h<\min_{j=1,\ldots,r}a_j.
\]
Since $K$ is a cone containing zero, its finite supremum is zero. Hence every $a_j>0$, and $a\cdot(U(w)-U(z))\leq0$ for every $w\in C$.

Conversely, every maximizer of a strictly positive weighted utility sum
is Pareto optimal. The Pareto optimal set is therefore the union of the
exposed faces of $C$ obtained by maximizing such sums. A polytope has
finitely many faces, so this union is closed.
\end{proof}

\begin{lemma}\label{lem:SWT}
Every Pareto optimal mass vector in $P(b)$ can be supported as a Lindahl equilibrium in the sense of Definition~\ref{def:lindahl-mass}. 

Every Pareto optimal random feasible allocation $\pi\in\Pi_{N,B}$ can be supported as a Lindahl equilibrium in the sense of Definition~\ref{def:lindahl-finite}.
\end{lemma}

\begin{proof}
Let $y^*\in P(b)$ be Pareto optimal. Multiplying a type's expected utility by its fixed positive mass does not change Pareto comparisons.
Apply Lemma~\ref{lem:positive-support} to the linear functions
$d_\theta(u_\theta\cdot x_\theta)=u_\theta\cdot y_\theta$.
There exist $a_\theta>0$ such that $y^*$ maximizes
$\sum_\theta d_\theta a_\theta(u_\theta\cdot x_\theta)$ on $P(b)$.
Define personalized price vectors by
\[
(p_\theta)_{\kappa s}
=\one\{\kappa=\theta\}\frac{a_\theta u_\theta(s)}{d_\theta},
\qquad w_\theta=p_\theta\cdot y^*.
\]
These prices and incomes are nonnegative, and
$p_\theta\cdot y=a_\theta(u_\theta\cdot x_\theta)$.
Thus a strict utility improvement over $y^*$ costs more than
$w_\theta$. The allocation $y^*$ maximizes utility in the budget set,
and every equally good allocation has the same expenditure
$w_\theta$. Finally,
\[
\left(\sum_\theta d_\theta p_\theta\right)\cdot y
=\sum_\theta d_\theta a_\theta(u_\theta\cdot x_\theta),
\]
which is maximized at $y^*$. This proves all the conditions of
Definition~\ref{def:lindahl-mass}.

For the finite-lottery assertion, apply
Lemma~\ref{lem:positive-support} to the simplex $\Pi_{N,B}$ and the linear
utilities $\mathbb E_\pi[u_{\theta_i}(\tilde s_i)]$. If $\pi^*$ is
Pareto optimal, it maximizes their weighted sum for some
$\alpha_i>0$. For each feasible deterministic allocation, set $p_i(s_1,\ldots,s_N)=\alpha_i u_{\theta_i}(s_i)$ and $w_i=\alpha_i\mathbb E_{\pi^*}[u_{\theta_i}(\tilde s_i)]$. 

Then, $p_i\cdot\pi=\alpha_i\mathbb E_\pi[u_{\theta_i}(\tilde s_i)]$
implies that $\pi^*$ maximizes utility in each budget set and minimizes
expenditure among the maximizers. Aggregate revenue is the positive
weighted utility sum, which is maximized at $\pi^*$. These are the
conditions of Definition~\ref{def:lindahl-finite}.
\end{proof}

\subsubsection{Proof of Theorem~\ref{thm:existence_lindahl}}

Fix $0<\epsilon<M$, and choose $\eta>0$ such that
\[
\eta<\frac1q\left(\frac\epsilon M\right)^{q-1}.
\] In particular, $\eta<1/q$. Define the $\eta$-simplex
\[ \Delta_\eta=\{\mu\in\Delta^{q-1}:\mu_\theta\geq\eta \text{ for every }\theta\in\Theta\}.
\] We interpret the elements of $\Delta_\eta$ as welfare weights. For $\mu\in\Delta_\eta$, let
\[
\Phi(\mu)=\argmax_{y\in Q}
\sum_{\theta\in\Theta}\mu_\theta u_\theta\cdot y_\theta
\]
be the set of maximizers of a weighted utilitarian welfare objective, when the weights are given by $\mu$. For $y\in Q$, with $x_\theta=y_\theta/d_\theta$, define
\[
E_\theta(y)=\sum_{\tau\in\Theta}
\max\{0,u_\theta\cdot x_\tau-u_\theta\cdot x_\theta-\epsilon\},
\] 
which may be interpreted as the total excess envy of type $\theta$ above the threshold $\epsilon$ at $y$.
Let
\[
\Gamma(y)=\argmax_{\nu\in\Delta_\eta}
\sum_{\theta\in\Theta}\nu_\theta E_\theta(y)
\] be the set of welfare weights when we try to give high weights to types that experience large amounts of envy. 

Define $\Psi:\Delta_\eta\times Q\rightrightarrows\Delta_\eta\times Q$
by $\Psi(\mu,y)=\Gamma(y)\times\Phi(\mu)$. Both objectives are
jointly continuous in their parameters and choice variables and linear
in the choice variables. Their domains are nonempty compact convex
sets. Hence, $\Psi$ has nonempty compact convex values and is upper
hemicontinuous. Kakutani's fixed point theorem gives
$(\mu,y)\in\Psi(\mu,y)$.

Because $y\in\Phi(\mu)$ and every $\mu_\theta\geq\eta>0$, the
allocation $y$ is Pareto optimal. Its welfare objective can also be
written as $\sum_\theta d_\theta\mu_\theta(u_\theta\cdot x_\theta)$.

We show that $y$ is $\epsilon$-envy-free. Suppose otherwise, so $\max_\theta E_\theta(y)>0$. Since $\mu\in\Gamma(y)$, every type whose $E_\theta(y)$ is below this maximum has weight $\eta$. Consequently, there is a maximizing type $\theta_0$ with $\mu_{\theta_0}\geq1/q$.

First observe that the directed graph of $\epsilon$-envy contains no cycles. If $(\theta_1,\ldots,\theta_L)$ were such a cycle, Lemma~\ref{lem:cycle} would give a feasible exchange of some size $t>0$. Its welfare gain would be
\[
t\sum_{j=1}^L\mu_{\theta_j}
u_{\theta_j}\cdot(x_{\theta_{j+1}}-x_{\theta_j})
>tL\eta\epsilon>0,
\]
contradicting $y\in\Phi(\mu)$.

By acyclicity, a directed path from $\theta_0$ reaches a sink
$\theta_L$ that $\epsilon$-envies nobody, in at most $q-1$ steps.
At the sink, $E_{\theta_L}(y)=0$, so
$\mu_{\theta_L}=\eta$.

For any edge $\kappa\to\tau$ along this path, use
Lemma~\ref{lem:cycle} for the two-type cycle. Welfare optimality gives
\[
\mu_\kappa u_\kappa\cdot(x_\tau-x_\kappa)
+\mu_\tau u_\tau\cdot(x_\kappa-x_\tau)\leq0.
\]
The first utility difference exceeds $\epsilon$, and the second is at
least $-M$. Therefore
$\mu_\kappa\epsilon-\mu_\tau M<0$, or
$\mu_\kappa<(M/\epsilon)\mu_\tau$. Tracing this inequality backwards along the path, and using $\mu_{\theta_L}=\eta$, gives
\[
\frac1q\leq\mu_{\theta_0}
<\eta\left(\frac M\epsilon\right)^L
\leq\eta\left(\frac M\epsilon\right)^{q-1}
<\frac1q,
\]
a contradiction. Thus $y$ is $\epsilon$-envy-free.

Now take $\epsilon_\ell=M/(\ell+1)$ for each $\ell=1,2,\ldots$, and select one such allocation $y^\ell$.
By compactness of $Q$, a subsequence
converges to some $y^*\in Q$. The marginal lotteries depend
continuously on $y$, so the $\epsilon_\ell$-envy inequalities imply
exact envy-freeness at $y^*$. Each $y^\ell$ is Pareto optimal, and since Lemma~\ref{lem:positive-support} shows that the set of Pareto optimal points is closed, their limit $y^*$ is Pareto optimal.
Apply that lemma again to obtain positive weights
$a_\theta$ supporting $y^*$ for the objective in
\eqref{eq:weighted-welfare}. These weights are selected at the limit;
they need not be limits of the auxiliary weights $\mu$.
When $Q=P(b)$, Lemma~\ref{lem:SWT} supplies the Lindahl support of $y^*$.

When $Q=P_N(b)$, implement $y^*$ by Lemma~\ref{lem:implementation}, obtaining $\pi^*\in\Pi_{N,B}$ with common marginals $x_\theta^*$ within each type. It is therefore envy-free. For any feasible lottery $\pi$, its induced mass vector satisfies $y^\pi\in P_N(b)$, and
\[
N\sum_{\theta\in\Theta}\sum_{s\in\bun} a_\theta u_\theta(s)y^\pi_{\theta s} =\sum_{i=1}^N a_{\theta_i} \mathbb E_\pi[u_{\theta_i}(\tilde s_i)].
\]
This identity holds for every feasible lottery $\pi$, whether or not agents of the same type receive the same marginal lottery. Since $y^*$ maximizes the left-hand side, $\pi^*$ maximizes a strictly positive weighted sum of all individual utilities over $\Pi_{N,B}$. It is Pareto optimal against all feasible lotteries. The construction in Lemma~\ref{lem:SWT}, with $\alpha_i=a_{\theta_i}$, gives its Lindahl support. That lemma also proves the assertion for every Pareto optimal $\pi\in\Pi_{N,B}$.\qed

\subsection{Sensitivity analysis}\label{sec:sensitivity}

For incentive compatibility in the large economy to translate into approximate incentive compatibility in a finite economy, we need to control how the large economy outcomes depend on certain parameters. Here we develop the needed ``sensitivity'' arguments. The first lemma is the Euclidean form of the error bound in \citet{Hoffman1952}. The integer proximity lemma is a standard-form specialization of the phenomenon proved in \citet[Theorem 1]{CookEtAl1986}; a direct proof is included so no rationality restriction on utility coefficients is required.

\medskip

Lemma~\ref{lem:hoffman} provides a uniform linear error bound. It will be used in Lemma~\ref{lem:optimal-sets} to control how the feasible set $P(b)$, and subsequently its optimal face $F(b)$, change when the parameter $b$ changes.

\begin{lemma}\label{lem:hoffman}
Let $G$ be a real matrix. There is a constant $H_G$ such that, whenever $C(h)=\{x:Gx\leq h\}$ is nonempty,
\[
\dist(v,C(h))\leq H_G\norm{(Gv-h)_+},
\qquad \dist(v,C)=\inf_{w\in C}\norm{v-w}.
\]
The constant does not depend on $h$ or $v$. Any constant $H_G$
satisfying this bound is called a Hoffman constant for $G$.
\end{lemma}

\begin{proof}
Let $w$ be the projection of $v$ onto $C(h)$. Write $g=v-w$. If $g=0$ there is nothing to prove.  Since $w$ is a projection, we have $g\cdot (z-w)\leq 0$ for all $z\in C(h)$. 
Write $G\in\mathbb R^{m\times k}$, let $G_j$ denote its $j$-th row, and define the active set at $w$ by $J=\{j:G_jw=h_j\}$. Consider a direction $d\in \R^k$ with $G_J d\leq 0$. For all sufficiently small $t>0$, $w+td\in C(h)$. Hence $g\cdot d\leq 0$ whenever $G_J d\leq 0$.

Consider the following two linear systems:
\begin{align*}
g & = G_J^{\mathsf T}\lambda, \quad \lambda\geq0;  \\
G_J d & \leq 0, \quad g\cdot d > 0 \text{ for some } d.
\end{align*}
By Farkas' lemma, exactly one of these systems has a solution. The second system has no solution by the previous argument. Conclude that there exists a nonnegative vector $\lambda$ such that $g = G_J^{\mathsf T}\lambda$. 

Among all nonnegative representations of $g$ as a conic combination of active row normals, choose one with minimal support $I\subseteq J$. We may then write $g=G_I^{\mathsf T}\lambda$ with $\lambda_j>0$ for all $j\in I$. It is now easy to see that the set $\{G_j:j\in I\}$ is linearly independent. If it were not, there would exist a nontrivial linear combination of the $G_j$'s that equals zero. This would allow us to adjust the coefficients $\lambda_j$ while preserving both the representation of $g$ and nonnegativity, until one coefficient becomes zero, contradicting the minimality of the support. Therefore, we can always find a representation of $g$ as a nonnegative combination of linearly independent active row normals.

Now, because $G_I$ has linearly independent rows, $G_I G_I^{\mathsf T}$ is positive definite (it is automatically positive semidefinite and full row rank implies positive definiteness). Hence,
\[
\norm{g}^2
=\lambda^{\mathsf T}G_I G_I^{\mathsf T}\lambda
\geq
\sigma_{\min}(G_I G_I^{\mathsf T})\norm{\lambda}^2,
\]
where $\sigma_{\min}(G_I G_I^{\mathsf T})>0$ denotes the smallest eigenvalue of $G_I G_I^{\mathsf T}$, and the inequality follows from the standard inequality for the Rayleigh quotient.
Therefore,
\[
\norm{\lambda}
\leq
\frac{\norm{g}}
{\sqrt{\sigma_{\min}(G_I G_I^{\mathsf T})}}.
\]

There are finitely many independent row subsets, so $\norm{\lambda}\leq H_G\norm{g}$ for one common scalar $H_G>0$, obtained from the inverse square root of  their smallest eigenvalues. Since $G_Iw=h_I$ and $\lambda\geq0$, Cauchy--Schwarz and the previous bound on $\norm{\lambda}$ give
\[
\norm{g}^2
=\lambda^{\mathsf T}(G_Iv-h_I)
\leq \lambda^{\mathsf T}(G_Iv-h_I)_+
\leq \norm{\lambda}\norm{(Gv-h)_+}
\leq H_G\norm{g}\norm{(Gv-h)_+}.
\]
Since $g\neq0$, dividing by $\norm{g}$ and using
$\norm{g}=\dist(v,C(h))$ gives
\[
\dist(v,C(h))\leq H_G\norm{(Gv-h)_+},
\]
as required.
\end{proof}

Lemma~\ref{lem:optimal-sets} will later be applied in the proof of Theorem~\ref{thm:main} with $b'=b^*$. Since $y^*\in F(b^*)$, it implies that whenever $b$ is close to $b^*$, the optimal face $F(b)$ contains a point close to the benchmark $y^*$.

\begin{lemma}\label{lem:optimal-sets}
Fix $A$ and $c$, and let $ F(b)=\argmax_{y\in P(b)}c\cdot y$. Let $b$ and $b'$ be such that $P(b)$ and $P(b')$ are nonempty and compact. There is a constant $H=H_{A,c}$, independent of $b$ and $b'$, such that
\begin{equation*}
 d_H(F(b),F(b'))\leq H\norm{b-b'},
\end{equation*}
where $d_H$ is the Euclidean Hausdorff distance, defined for nonempty compact sets $K,L$ by
$d_H(K,L) =
\max\left\{
\sup_{x\in K}\dist(x,L),
\sup_{y\in L}\dist(y,K)
\right\}.$
\end{lemma}

\begin{proof}
For $y\in P(b)$, applying Lemma~\ref{lem:hoffman} to
$G=(A^{\mathsf T},-A^{\mathsf T},-I)^{\mathsf T}$ and
$h(b)=(b,-b,0)^{\mathsf T}$ gives
\[
\dist(y,P(b'))\le H_A\|(Gy-h(b'))_+\|=H_A\|b-b'\|,
\]
where $H_A$ depends only on $A$. By symmetry, the same bound holds with $b$ and $b'$ reversed. Hence $d_H(P(b),P(b'))\le H_A\|b-b'\|$.

Let $v(b)=\max_{y\in P(b)}c\cdot y$ be the optimal value.
Then,
\[
|v(b)-v(b')|
\leq \norm{c}\,d_H(P(b),P(b'))
\leq \norm{c}H_A\norm{b-b'}.
\]

Now write the optimal set as $F(b) = \{y:Ay=b,\ y\geq0,\ c\cdot y=v(b)\}$. Equivalently,
\[
F(b)=\{y:\widetilde G y\leq \widetilde h(b)\},
\] where
\[
\widetilde G = \begin{pmatrix}
 A \\
 -A \\
 -I \\
 c^{\mathsf T} \\
 -c^{\mathsf T}
\end{pmatrix},
\qquad \widetilde h(b) = \begin{pmatrix}
 b \\
 -b \\
 0 \\
 v(b) \\
 -v(b)
\end{pmatrix}.
\]
The matrix $\widetilde G$ depends only on $A$ and $c$. 
Let $H_{\widetilde G}$ be a Hoffman constant from Lemma~\ref{lem:hoffman} for $\widetilde G$.
Fix $y\in F(b)$.  Its constraint-violation vector relative to the system defining $F(b')$ is
\[
\bigl(\widetilde G y-\widetilde h(b')\bigr)_+ =
\begin{pmatrix}
 (b-b')_+\\
 (b'-b)_+\\
 0\\
 (v(b)-v(b'))_+\\
 (v(b')-v(b))_+
\end{pmatrix}.
\]
Since, for every vector $z$, $\|z_+\|^2+\|(-z)_+\|^2=\|z\|^2$, we obtain
\[ \left\|
\bigl(\widetilde G y-\widetilde h(b')\bigr)_+
\right\|^2 = \|b-b'\|^2+|v(b)-v(b')|^2. \]
Lemma~\ref{lem:hoffman}, together with the Lipschitz bound on the optimal value function $v(b)$ provides
\[
\begin{aligned}
\dist(y,F(b'))
&\leq
H_{\widetilde G}
\sqrt{\|b-b'\|^2+|v(b)-v(b')|^2}\\
&\leq
H_{\widetilde G}\sqrt{1+\|c\|^2H_A^2}\,
\|b-b'\|.
\end{aligned}
\]
Taking the supremum over $y\in F(b)$ and then repeating the same argument with $b$ and $b'$ interchanged gives 
\[
d_H(F(b),F(b')) \leq \left( H_{\widetilde G}\sqrt{1+\|c\|^2H_A^2} \right) \|b-b'\|.
\]
\end{proof}

Lemma~\ref{lem:proximity} will later be used to pass from the optimal face $F(b)$ of the continuous problem to the optimal face $F_N(b)$ of the finite economy. Together with Lemma~\ref{lem:optimal-sets}, it implies that when $b$ is close to $b^*$ and $N$ is large, $F_N(b)$ contains a point close to the benchmark $y^*$.

\begin{lemma}\label{lem:proximity}
Fix the same $A$ and $c$ as in Lemma~\ref{lem:optimal-sets}. There exists a scalar $\Lambda_A$ such that, for every $N$ and every $b$ such that $P(b)$ is nonempty and compact and $\mathcal Z_N(b)\neq\varnothing$,
\[
d_H(F_N(b),F(b))\leq \frac{\Lambda_A}{N}.
\]
The constant $\Lambda_A$ depends only on $A$ and is independent of $N$, $b$, and $c$.
\end{lemma}

\begin{proof}
For each coordinate sign pattern $\{-1,1\}^{qm}$, intersect $\ker A$ with the corresponding
closed orthant. This is a pointed rational polyhedral cone. Its extreme rays have integer generators. Taking their union over the finitely many sign patterns gives a finite set $\mathcal G_A\subset\Z^{qm}\setminus\{0\}$. By conic Carath\'eodory, every nonzero $h\in\ker A$ can be written as
\begin{equation}\label{eq:conformal}
 h=\sum_{j=1}^{\ell}\lambda_j g^j,\qquad  1\leq\ell\leq mq,\quad \lambda_j>0,\quad g^j\in\mathcal G_A.
\end{equation}
Since all the $g^j$ lie in the same closed orthant as $h$ and the coefficients are positive, there is no coordinatewise cancellation; hence $h_i=0$ implies $g_i^j=0$ for every $j$. Thus every summand is conformal to $h$, meaning that $g_i^jh_i\geq0$ and $g_i^j=0$ whenever $h_i=0$. If $\ker A=\{0\}$, let $\Lambda_A=0$. Otherwise, set $\Lambda_A=mq\max_{g\in\mathcal G_A}\norm{g}.$

Let $x$ solve $\max\{c\cdot x \colon Ax = Nb,\ x \geq 0\}$, and choose a solution $z$ to  $\max\{c\cdot z:Az=Nb,\ z\in\mathbb Z_+^{qm}\}$ that minimizes $\|z-x\|_1$ among the solutions. Note that $x/N\in F(b)$. If $h=z-x \neq 0$, use \eqref{eq:conformal}. For each $j$ and $0\leq s\leq\lambda_j$, both $x+sg^j$ and $z-sg^j$ are nonnegative and satisfy the same equalities. This follows coordinate by coordinate from conformality and $\lambda_j|g_i^j|\leq|h_i|$. The optimality of $x$ implies $c\cdot g^j\leq0$. If $\lambda_j\geq1$, the vector $z-g^j$ is integer feasible. Integer optimality then implies $c\cdot g^j\geq0$, so it is also optimal. It is strictly closer to $x$ in $\ell^1$, a contradiction. Thus every $\lambda_j<1$ and $\norm{z-x}\leq \Lambda_A$.

For the reverse direction, fix an integer optimum $z$ and choose a continuous optimum $x$ minimizing $\|z-x\|_1$. Let $h=z-x$. If $h=0$, the bound is immediate. Otherwise apply \eqref{eq:conformal} to this $h$, obtaining generators $g^j$ and coefficients $\lambda_j$. Note that  $g_i^jh_i \geq 0$, with $g_i^j = 0$ whenever $h_i = 0$. Since $x$ is optimal, $c\cdot g^j \leq 0$. 
We cannot have $c\cdot g^j=0$. Indeed, for sufficiently small $\varepsilon\in(0,\lambda_j]$, conformality implies that $x+\varepsilon g^j$ is feasible and lies strictly closer to $z$ in $\ell^1$. If $c\cdot g^j=0$, it would also be optimal, contradicting the choice of $x$. Hence $c\cdot g^j<0$. If any $\lambda_j \geq 1$, the integer feasible point $z-g^j$ would have a strictly higher objective. Thus again every $\lambda_j<1$ and the same bound follows.

Dividing by $N$, the first argument bounds the distance of every point of $F(b)$ to $F_N(b)$ by $\Lambda_A/N$. The second bounds the distance of every normalized integer optimum to $F(b)$ by the same amount. 
Since $F(b)$ is convex, the same bound extends from normalized integer optima to their convex hull, namely $F_N(b)$. Together, these bounds give the asserted Hausdorff estimate.
\end{proof}

\begin{lemma}\label{lem:projection}
Let $K,L\subseteq\Delta^{D-1}$ be nonempty compact convex sets and $z\in\Delta^{D-1}$. If $p=\operatorname{proj}_Kz$ and $q=\operatorname{proj}_Lz$, then
\begin{equation*}
 \norm{p-q} \leq 2\sqrt{d_H(K,L)}.
\end{equation*}
\end{lemma}
\begin{proof}
Let $h=d_H(K,L)$. Choose $q'\in K$ and $p'\in L$ with $\norm{q-q'}\leq h$ and $\norm{p-p'}\leq h$. 
By the characterization of Euclidean projections,
$\ip{z-p}{q'-p}\leq 0$ and $\ip{z-q}{p'-q}\leq 0$.
Hence
\[
 \ip{z-p}{q-p}\leq\norm{z-p}h,\qquad
 \ip{z-q}{p-q}\leq\norm{z-q}h.
\]
Adding the two inequalities yields $\norm{p-q}^2 \leq (\norm{z-p}+\norm{z-q})h \leq 4h$, as all three points lie in the unit simplex.
\end{proof}

\subsection{Proof of Theorem~\ref{thm:main}} \label{sec:pfmain}

Fix the pair $(y^*,a)$ used to construct the mechanism. For the remainder of this proof, write $c_\theta=a_\theta u_\theta$ so that 
\[ c\cdot y =\sum_{\theta\in\Theta}\sum_{s\in\bun} a_\theta u_\theta(s)y_{\theta s},
\]
and the choice of $(y^*,a)$ implies $y^*\in F(b^*)$. At a report profile $r$, the mechanism is given by \eqref{eq:zopt}--\eqref{eq:mechanism} with $b=b(r)$.

\begin{theorem}\label{thm:main-quantitative}
  Fix $b^*=(d^*,\sigma^*)$ with $d^*\gg0$ and $P(b^*)\neq\varnothing$.  Choose $(y^*,a)$ by Theorem~\ref{thm:existence_lindahl} applied with $Q=P(b^*)$, and let $m_{N,B}$ be the mechanism constructed above using this fixed choice.

There exist constants $C>0$, $\rho>0$, and $N_0\in\mathbb N$, independent of $N$, $B$, and the report profile, with the following properties for every feasible pair $(N,B)$.

\begin{enumerate}
\item
At every report profile $r\in\Theta^N$, the random allocation $\pi=m_{N,B}(r)$ is supported on feasible deterministic allocations and gives identical reports identical marginal lotteries. It is Pareto optimal with respect to the utilities $(u_{r_i})_{i=1}^N$, with comparisons taken over all random feasible allocations, including those that do not treat identical reports equally. In particular, its outcome is Pareto optimal at every truthful profile.

\item
Suppose $N\geq N_0$, and fix a report profile $r$ such that $ b=b(r)\coloneqq (d(r),B/N)$ and $e\coloneqq \|b-b^*\|\leq\rho$.   Write $\pi=m_{N,B}(r)$. For every pair of agents $i,j$,
\[
 \mathbb E_{\pi}[u_{r_i}(\tilde s_i)]  \geq  \mathbb E_{\pi}[u_{r_i}(\tilde s_j)]  -C\left(e+\frac1N\right).
\]
For every agent $i$ and every alternative report $\tau\in\Theta$,
write $\pi'=m_{N,B}(\tau,r_{-i})$. Then
\[
 \mathbb E_{\pi}[u_{r_i}(\tilde s_i)]  \geq  \mathbb E_{\pi'}[u_{r_i}(\tilde s_i)]  -C\left(e+\frac1N\right).
\]
The incentive inequality applies to an agent whose true type is $r_i$, uniformly over opponents' reports at such baseline profiles. It also holds for any randomized alternative report. Consequently, at every type profile $\theta$ satisfying these conditions, the mechanism is both $C(\|b-b^*\|+N^{-1})$-envy-free and $C(\|b-b^*\|+N^{-1})$-incentive compatible.

\item
Under the conditions of item~2, consider any agent $i$ and any alternative report $\tau$, and set $\widehat b=b(\tau,r_{-i})$. For a type $\kappa$ with positive share, the common marginal assignments at the two profiles are
\[
 x_\kappa^N(b)=\frac{y_\kappa^N(b)}{d_\kappa(r)},
 \qquad
 x_\kappa^N(\widehat b)  =\frac{y_\kappa^N(\widehat b)}{d_\kappa(\tau,r_{-i})}.
\]
The constants $\rho$ and $N_0$ can be chosen so that every type has positive share both at $r$ and after every unilateral report
change. For all such changes,
\begin{equation}\label{eq:impact}
 \max_{\kappa\in\Theta}  \bigl\|x_\kappa^N(\widehat b)-x_\kappa^N(b)\bigr\|  \leq \frac{C}{\sqrt N}.
\end{equation}
\end{enumerate}
\end{theorem}

To prove Theorem~\ref{thm:main} from Theorem~\ref{thm:main-quantitative}, fix $\epsilon>0$ and choose
\[
 \eta=\min\left\{\rho,\frac{\epsilon}{2C}\right\},  \qquad  N_\epsilon=\max\left\{N_0,
          \left\lceil\frac{2C}{\epsilon}\right\rceil\right\}.
\]
If $N\geq N_\epsilon$ and $\|b-b^*\|\leq\eta$, then $C(\|b-b^*\|+N^{-1})\leq\epsilon$, so item~2 gives $\epsilon$-envy-freeness and $\epsilon$-incentive compatibility. Item~1 gives exact feasibility, equal treatment of identical reports, and Pareto optimality at every report profile.

\paragraph{Pareto optimality:}

Fix a report profile $r=(r_1,\ldots,r_N)$ and let $y^N=y^N(b)$ denote the mass vector selected by the mechanism.  Recall that $c_{\theta s}=a_\theta u_\theta(s)$. Since $F_N(b) = \conv\{z/N:z\in\mathcal Z_N^{\rm opt}(b)\},$ and every $z\in\mathcal Z_N^{\rm opt}(b)$ maximizes $c\cdot z$ over
$\mathcal Z_N(b)$, the linearity of the objective implies that every $y\in F_N(b)$ maximizes $c\cdot y$ over $P_N(b)$.

For any feasible lottery $\pi$ over deterministic allocations, define its induced normalized expected count vector by $y^\pi_{\theta s} = \frac1N \sum_{i:r_i=\theta}\Pr_\pi(\tilde s_i=s).$ Every deterministic allocation in the support of $\pi$ has an integer count table in $\mathcal Z_N(b)$. Taking expectations, therefore, gives $y^\pi\in P_N(b).$ Moreover,
\begin{align*}
N c\cdot y^\pi &= N\sum_{\theta,s}a_\theta u_\theta(s)y^\pi_{\theta s} = \sum_{\theta,s}a_\theta u_\theta(s) \sum_{i:r_i=\theta}\Pr_\pi(\tilde s_i=s) \\
&= \sum_{i=1}^N a_{r_i}\sum_s u_{r_i}(s)\Pr_\pi(\tilde s_i=s) = \sum_{i=1}^N a_{r_i}\, \mathbb E_\pi[u_{r_i}(\tilde s_i)]. 
\end{align*}

Suppose, toward a contradiction, that some feasible lottery $\pi$ Pareto dominates the mechanism's lottery with respect to reported utilities. Then, since $a_{r_i}>0$ for every $i$, it follows that
\[
\sum_{i=1}^N a_{r_i}\, \mathbb E_\pi[u_{r_i}(\tilde s_i)] > \sum_{i=1}^N a_{r_i}\, \mathbb E[u_{r_i}(\tilde s_i)].
\]
This is equivalent to $N c\cdot y^\pi>N c\cdot y^N.$ But $y^\pi\in P_N(b)$, which yields a contradiction. Therefore, no feasible lottery Pareto dominates the mechanism's lottery.

\paragraph{Approximate envy:}
Write $H=H_{A,c}$ and $\Lambda=\Lambda_A$. Lemmas~\ref{lem:optimal-sets} and~\ref{lem:proximity} imply that 
\begin{equation*}
 \norm{y^N(b)-y^*}  =\dist(y^*,F_N(b)) \leq H\norm{b-b^*}+\Lambda/N.
\end{equation*}
Indeed, first find an optimum in $F(b)$ within $H\norm{b-b^*}$ of $y^*$, and then an integer optimum within $\frac{1}{N}\Lambda$ of that point.

Let $d_{\min}^*=\min_\theta d_\theta^*>0$, set $\alpha=d_{\min}^*/2$ and $\rho=d_{\min}^*/4$, and take $N_0\geq4/d_{\min}^*$. For $e=\norm{b-b^*}\leq\rho$ and $N\geq N_0$, every type share at the baseline and at any unilateral deviation is at least $\alpha$. Since $\norm{y_\theta^*}\leq d_\theta^*$,
\begin{equation}
 \norm{x_\theta^N(b)-x_\theta^*}  \leq\frac{\norm{y_\theta^N(b)-y_\theta^*} +|d_\theta-d_\theta^*|}{d_\theta}
 \leq\frac{(H+1)e+\Lambda/N}{\alpha}.
 \label{eq:percap-target}
\end{equation}
Let $U=\max_\theta\norm{u_\theta}\leq M\sqrt m$. Envy-freeness of $x^*$ now gives, for every pair $\theta,\tau$,
\begin{equation}\label{eq:envy-explicit}
 \pos{u_\theta\cdot x_\tau^N(b)-u_\theta\cdot x_\theta^N(b)}  \leq\frac{2U}{\alpha}\bigl((H+1)e+\Lambda/N\bigr).
\end{equation}

\paragraph{Incentives:}
A type-$\theta$ agent reporting type $\tau$ changes $b=(d,\sigma)$ into $\widehat b$ by  $\widehat b-b=((\one_\tau-\one_\theta)/N,0)$. So $\norm{\widehat b-b}\leq\sqrt2/N$ and $\norm{\widehat b-b^*}\leq e+\sqrt2/N$. The gain from reporting $\tau$ is
\begin{align*}
 u_\theta\cdot x_\tau^N(\widehat b)-u_\theta\cdot x_\theta^N(b) ={}&u_\theta\cdot(x_\tau^*-x_\theta^*) \\
 &+u_\theta\cdot(x_\tau^N(\widehat b)-x_\tau^*)  -u_\theta\cdot(x_\theta^N(b)-x_\theta^*). 
\end{align*}
The first term is nonpositive. Since every type share is at least $\alpha$ at both profiles, the bound in \eqref{eq:percap-target} applies to both $b$ and $\widehat b$, yielding the uniform bound
\begin{equation}
u_\theta\cdot x_\tau^N(\widehat b) -u_\theta\cdot x_\theta^N(b) \leq \frac{U}{\alpha} \left( 2(H+1)\norm{b-b^*}+ \frac{2\Lambda+\sqrt2(H+1)}{N}
\right). \label{eq:incentive-explicit}
\end{equation}
Only this agent's true label and the opponents' reports enter the argument.
Yes let us 
\paragraph{Uniform effect of one report}

Our sensitivity analysis yields
\[
 d_H(F_N(b),F_N(\widehat b)) \leq 2\Lambda/N+H\norm{b-\widehat b}.
\]

If one agent misreports $\tau$ in place of $\theta$, $d$ changes to $\widehat d$ and $b$ to $\widehat b=(\widehat d,\sigma)$ with $\|\widehat b-b\|\leq\frac{\sqrt2}{N}$. Apply Lemma~\ref{lem:projection} with $z=y^*$, $p=y^N(b)$, and $q=y^N(\widehat b)$ to obtain:
\[
 \norm{y^N(\widehat b)-y^N(b)} \leq 2\sqrt{d_H(F_N(b),F_N(\widehat b)) } \leq 2\sqrt{(2\Lambda+\sqrt2H)/N}.
\]

For any type $\kappa$, we obtain a marginal assignment from the corresponding mass vector by dividing by its weight in the distribution; these are bounded below, which allows us to use the previous estimates on the mass vectors. Specifically, $x^N(\widehat b)_\kappa = y^N(\widehat b)_\kappa/\widehat{d}_\kappa$ and $x^N(b)_\kappa = y^N(b)_\kappa/d_\kappa$. So, 
\[
x_\kappa^N(\widehat b)-x_\kappa^N(b)
={} \frac{y_\kappa^N(\widehat b)-y_\kappa^N(b)}
     {\widehat d_\kappa} +\frac{d_\kappa-\widehat d_\kappa}{\widehat d_\kappa}
\,x_\kappa^N(b).
\]
A single report changes any type's share by at most $1/N$, and $\widehat d_\kappa\geq\alpha$. Since $x^N_\kappa(b)$ is a probability vector, $\norm{x^N_\kappa(b)}\leq1$. Hence, 
\begin{equation}\label{eq:impact-explicit}
 \norm{x^N_\kappa(\widehat b)-x^N_\kappa(b)}
 \leq\frac1\alpha\left(2\sqrt{\frac{2\Lambda+\sqrt2H}{N}}+\frac{1}{N}\right).
\end{equation}
This proves \eqref{eq:impact}.

Let the constant $C$ absorb the coefficients of \eqref{eq:envy-explicit}, \eqref{eq:incentive-explicit}, and \eqref{eq:impact-explicit}. \qed

\bibliographystyle{plainnat}
\bibliography{replica.bib}

\end{document}